\documentclass[aps,twocolumn,twoside,floatfix,nofootinbib,a4paper]{revtex4}
\usepackage{graphicx}
\usepackage{marginnote}

\usepackage{hyperref}
\usepackage{enumitem,kantlipsum}
\usepackage[usenames,dvipsnames]{color}
\usepackage{amsmath}
\usepackage{amssymb}
\usepackage{amsfonts}
\usepackage{mathrsfs}
\usepackage{bm}
\usepackage[all]{xy}
\usepackage{hyperref}
\usepackage{tikz-cd}

\newcommand{\beq}{\begin{equation}}
\newcommand{\eeq}{\end{equation}}
\newcommand{\bea}{\begin{eqnarray}}
\newcommand{\eea}{\end{eqnarray}}
	
\newcommand{\ea}{\end{align*}}

\newcommand{\bma}{\begin{pmatrix}}
\newcommand{\ema}{\end{pmatrix}}

\usepackage[percent]{overpic}
\usepackage{overpic}

\begin{document}
\title{Gravity-driven magnetogenesis}
\author{Fan Zhang} 
\affiliation{Gravitational Wave and Cosmology Laboratory, Department of Astronomy, Beijing Normal University, Beijing 100875, China}
\affiliation{Advanced Institute of Natural Sciences, Beijing Normal University at Zhuhai 519087, China}

\date{\today}

\begin{abstract}
\begin{center}
\begin{minipage}[c]{0.9\textwidth}
Structure formation heralds the era of deviation of the matter content of the Universe away from thermal equilibrium, so the gravitational contribution to entropy, in the form of Weyl curvature, must become active in order for the overall entropy of the Universe to remain increasing. The tidal and frame dragging sectors of the Weyl tensor must inevitably both be present in this dynamic environment, as they mutually induce each other. The frame dragging effect is able to impress vorticity onto the plasma current arising due to the mass disparity between electrons and protons, which in turn begets a magnetic field from none. We show that this gravity-driven magnetogenesis mechanism, besides being able to operate outside of galaxies, thus facilitate large coherence length scales, may be able to generate the field strength necessary to seed dynamo processes.  
 \end{minipage} 
 \end{center}
  \end{abstract}
\maketitle

\raggedbottom
\section{Introduction and overview \label{sec:Intro}} 
The existence of long range interactions like either gravity or magnetism, breaks additivity \cite{2006JPhCS..31...18B} and invalidates thermodynamics in its standard formulation. This is the reason why there can be cosmic structures forming out of initial thermal equilibrium. The gravitational contribution to this process is well appreciated, in the form of Weyl curvature tensor being speculated to provide a generalized sense of entropy \cite{1979grec.conf..581P}. Less recognized is that the gradual strengthening and fragmentation in the magnetic sector, also endowing a long range field with an increasing degree of configurational complexity, could perhaps also offer an analogous additional source of generalized entropy growth. Therefore the magnetization of the Universe, over time, may also be entropically favorable. 
In lockstep with gravity, it is plausible that more cosmological processes would produce weaker magnetic fields that are coherent on larger length scales to begin with, and later on, more localized astrophysical processes occurring within the formed structures, such as dynamos within galaxies (see e.g., \cite{2023ARA&A..61..561B}), stars and planets, would manufacture stronger fields that are more local, thereby achieving both strengthening and fragmentation over time. 

The cosmic scale fields would also serve as seeds to later dynamos, so ascertaining their properties would be important. Observationally, magnetic fields are indeed indirectly measured within intergalactic space [see e.g., \cite{2010Sci...328...73N,2013A&ARv..21...62D}; 
lower limit @ $\sim \mathcal{O}(10^{-18})$G to $\sim \mathcal{O}(10^{-14})$G \cite{2010MNRAS.406L..70T,2011A&A...529A.144T,2017ApJ...835..288A,2011ApJ...733L..21D,2012ApJ...747L..14V,2023arXiv231104273T}], which can be coherent on cosmological length scales of Mpc and above (see e.g., \cite{2023arXiv231104273T}). If they are of primordial origins, a major hurdle arises from the fact that the magnetic field strength dilutes in the same way as its electromagnetic cousin, light. Considering how light once reigned over the radiation dominated era, but is now only but a slight whimper in the form of the Cosmic Microwave Background, the primordially generated magnetic field, and the electric field involved in its generation, must have been exceedingly strong for it to leave a meaningful vestige today, but complications like the Swinger effects could set in to shut the magnetogenesis down. 

Consequently, it is worthwhile exploring also the possibility of processes occurring later in cosmic history, when the scale factor of the Universe is closer to what it is today, so the magnetic field can be born directly into small amplitudes. Processes beginning just after reionization would be particularly interesting, as this era is when electromagnetism (beyond light) re-awakens and active dynamics in the form of structure formation is ongoing. One then notices that this period is also when the gravitational entropy levitates, as inhomogeneity and anisotropy builds up, providing the necessary conditions for Weyl curvature to thrive. It is therefore interesting to query the possible interplay between the simultaneous germination of gravitational and magnetic entropies. Perhaps the timing coincidence is not accidental, and Weyl curvature in fact drives magnetogenesis in the first place. 

Fortuitously for this supposition, when one breaks the Weyl curvature down into its gravitoelectric ${\bf g}$ (same as Newtonian gravitational acceleration, see e.g., \cite{2009PhRvD..80l4014K,1961PIRE...49..892F,1986bhmp.book.....T}) and gravitomagnetic ${\bf H}$ (describing frame dragging effects; see e.g., \cite{1998MNRAS.295L...6K,2013PhPl...20b2901A,2020A&A...642L...6B} for some examples of its influence on magnetogenesis near black holes) components, its evolution equation reduces to a form analogous to Maxwell's equations, and the geodesic equation analogous to the Lorentz force expression. This means first of all, gravitomagnetism must be present, because structure formation is necessarily concomitant with an evolving gravitational field, thus $\partial_t {\bf g} \neq 0$, which then produces ${\bf H}$ via induction, just like a changing electric field will induce a magnetic field. Secondly, since ${\bf H}$ acts on the newly reionized plasma particles like a magnetic field (see Eq.~\ref{eq:momentum} below), with a cross product to their velocities, it will usher them into vortical motions that is necessary for magnetogenesis. In short, by applying  equations similar in form backwards and forwards, we have a close coupling between gravitomagnetism and magnetism, via a conduit provided by plasma motion. The sprouting gravitomagnetism during structure formation necessarily ignite magnetism though this coupling, the issue is how strong this effect is, or in other words how large a magnetic field can be created this way.  

\section{Semi-quantitative estimate}
The two fluid equations for the proton-electron plasma, in the presence of a gravitoelectromagnetic field, are ($j, l \in \{{\rm electron, proton}\}$; in SI units; see e.g., \cite{2016ippc.book.....C,2011PhRvD..84l4014N})
\bea
&&\frac{\partial n_j}{\partial t} + \nabla \cdot (n_j {\bf v}_j) = 0\,, \label{eq:continuity}\\
&& m_j n_j \left(\frac{\partial {\bf v}_j}{\partial t} + ({\bf v}_j\cdot \nabla) {\bf v}_j \right) = n_j q_j({\bf E}+ {\bf v}_j\times {\bf B})  \label{eq:momentum}\\ 
&&
\quad \quad - \nabla p_j + m_j n_j \left({\bf g} + {\bf v}_j \times {\bf H} \right) - \sum_l \bar{\nu}_{jl} n_j m_j({\bf v}_j-{\bf v}_l)\,, \notag \\
&& p_j = n_j k T_j\,, 
\eea 
and the Maxwell's equations 
\bea
\nabla \cdot {\bf E} = \frac{\rho}{\epsilon_0} \,,&\quad&
\nabla \times {\bf E} = -\frac{\partial B}{\partial t} \,,\notag \\
\nabla \cdot {\bf B} = 0\,, &\quad& \nabla \times {\bf B} = \mu_0 {\bf j} + \frac{1}{c^2}\frac{\partial {\bf E}}{\partial t} \,, \label{eq:Maxwell}
\eea
with sources $\rho = \sum_j q_j n_j$, ${\bf j} =\sum_j q_j n_j {\bf v}_j$, and $m_j$ are the masses, $n_j$ the number densities, ${\bf v}_j$ the velocities, $T_j$ the temperatures, $p_j$ the pressures, of the plasma species. 

As ${\bf g}$ drives the plasma towards the bottoms of gravitational potential wells, it tends to accelerate the plasma, heating it up, establishing a temperature gradient across different gravitational potential ``heights''. The resulting pressure gradient will affect the immobile ($m_p\gg m_e$) protons less, leading to a phase space differentiation\footnote{This depends on the difference in inertial masses. The presence of gravity, for which electrons and protons possess vastly different charges in the form of gravitational masses, also makes their differentiation possible even in force balanced non-accelerating situations, e.g., trying to balance ${\bf g}$ against pressure gradience in an isothermal setting, will result in barometric distributions in number density, with protons concentrating more towards the bottom of the gravitational potential well.} between the two charged species, and subsequently an awakening of the electromagnetic sector. Consequently, a conduit for energy transfer between the gravitational and electromagnetic sectors is opened up. In particular, a form of effective inductance of the plasma flow, produced by gravitomagnetism guiding the currents into vortical motions, becomes active, that tries to resist further increases of the current, by draining some of the injected gravitational energy into the magnetic fields. 

Quantitatively, we have that 
\bea
-{\bf E} \approx {\bf v}_j\times {\bf B}  - \frac{\nabla p_j}{n_j q_j} + \frac{m_j}{q_j} \left({\bf g} + {\bf v}_j \times {\bf H} \right) -\frac{m_j}{q_j} {\bf a}_j\,,
\eea 
where we have denoted the Lagrangian picture acceleration as ${\bf a}_j$ in the interest of brevity, and in our case with a low density plasma, we ignore the impact of collisions\footnote{This collisionless assumption does not remove the pressure gradient term, because it comes not from collision, but momentum exchange due to particles moving in and out of the Lagrangian picture volume element via thermal motion, which is why cross-species collisions does not enter into this term at all.} on magnetogensis by setting $\bar{\nu}_{jk}=0$. Hitting both sides with $\nabla \times $ to isolate the non-vanishing vorticities relevant for magnetogenesis, and use Eq.~\eqref{eq:Maxwell}, we obtain 
\begin{align} \label{eq:Btime1}
\frac{\partial {\bf B}}{\partial t} \approx& \nabla \times \left( {\bf v}_j\times {\bf B} \right) -\frac{\nabla n_j \times \nabla p_j}{q_j n_j^2} \notag \\
&+ \frac{m_j}{q_j} \left[\nabla \times ({\bf g} -{\bf a}_j)+ \nabla\times ({\bf v}_j \times {\bf H}) \right]\,.
\end{align}
The second term is the Biermann thermal battery. It is not what we are trying to examine in this note, so we ignore this term below (see e.g., \cite{1997ApJ...480..481K} for its effects). As well, the first term in Eq.~\eqref{eq:genesis} is the usual flux freezing term involved more with dynamo processes, so we also ignore it. In appropriate coordinates, ${\bf g}$ is the gradient of the squared lapse (see e.g., \cite{2011PhRvD..84l4014N}), so its curl vanishes. 

Some of the energy injected by the gravitational sector into the vortical motion of the plasma promptly moves on through into the magnetic sector. Some may however be intercepted and hoarded by the plasma itself, in the form of an increasing kinetic energy associated with vortical motion, and $\nabla \times {\bf a}_j$ gives an indication of how fast this buildup is occurring. On general grounds, we expect the kept fraction to be small, because electrons and protons have much larger charges than masses (they would be overcharged naked singularities if viewed as Kerr-Newman black holes), so whenever parity is broken between them, they are quite apt at creating large currents as compared to their kinetic momentum. These large currents act as wide floodgates, that the small amount of vortical motion kinetic energy, injected by gravity through a much narrower window (supplied to the plasma by ${\bf g}$ and channelled into the vortical motion compartment by ${\bf H}$), can quickly evacuate through. Therefore, we also ignore $\nabla \times {\bf a}_j$ from here on. 

On the other hand, ${\bf H}$ is the curl of the shift vector, so its divergence vanishes. The divergence of ${\bf v}_j$ is expected to be small in the absence of strong sources and sinks of plasma (may require further investigations if reionization is ongoing\footnote{Because reionization is unlikely directly connected with $\bf H$, we do not expect adding this influence would result in cancellations with the terms in Eq.~\eqref{eq:genesis}. In fact, if $\nabla \cdot {\bf v}_j$ is large enough, the term $(\nabla \cdot {\bf v}_j){\bf H}$ that we have ignored may become an additional significant source for magnetism.}), and when the plasma is incompressible, or more precisely not subjected to strong compressional influences. All together, we reduce Eq.~\eqref{eq:Btime1} to
\bea \label{eq:genesis}
\frac{\partial {\bf B}}{\partial t} \approx \frac{m_j}{q_j} \left[({\bf H}\cdot \nabla) {\bf v}_j - ({\bf v}_j\cdot \nabla) {\bf H} \right]\,,
\eea
which accommodates the possibility for the creation of ${\bf B}$ where there was none initially, i.e., magnetogenesis. 

Within these source terms provided by gravitomagnetism, the distributions of ${\bf v}_j$ and ${\bf H}$ both vary on the scale of the cosmic structures, thus the two terms are of similar sizes, yet should not be so fine tuned as to almost exactly cancel. This is because ${\bf H}$ enters into the acceleration of the charged particles, whose masses are non-negligible this late into cosmic history when temperatures are low. The particles thus have inertia, so their motions won't conform instantaneously onto the ${\bf H}$ settings at the same moment and location\footnote{Unlike how charges in a force-free plasma (involves extremely strong electromagnetic field, not the case for us) would in regard of ${\bf B}$.}, but instead also depend on past motion. So ${\bf v}_j$ won't synchronize with ${\bf H}$ in general, and the difference between the two terms in the square bracket of Eq.~\eqref{eq:genesis} would likely be of the same order of magnitude as that of each individual term. We can now make some further progress by noting that 
\bea
({\bf v}_j\cdot \nabla) {\bf H} = -2 {\bf v}_j \cdot \mathcal{B}\,,
\eea
where $\mathcal{B}$ is the vortex tensor \cite{2011PhRvL.106o1101O,2011PhRvD..84l4014N} (describing differential frame dragging, so it is the gradient of ${\bf H}$), and the dot product denotes contraction with its first index. We now end up with [leaving out unimportant $\mathcal{O}(1)$ numerical factors]
\bea \label{eq:genesisorder}
\frac{\partial {\bf B}}{\partial t} \sim \frac{m_j}{q_j} {\bf v}_j \cdot \mathcal{B} \,.
\eea

In order to arrive at a (crude) estimate of the field strength predicted by Eq.~\eqref{eq:genesisorder}, we would need an estimate on the size of $\mathcal{B}$. To this end, we note that the vortex tensor $\mathcal{B}$ and the tendex tensor $\mathcal{E}$ \cite{2011PhRvL.106o1101O,2011PhRvD..84l4014N} (is the double spatial derivative of the gravitational potential $\Phi$, or minus the gradient of ${\bf g}$, so it measures differential acceleration resulting from tidal influences) mutually induce each other in a form mirroring that of Eq.~\eqref{eq:Maxwell}, which in near vacuum (in dense matter, Ricci half of the Riemann tensor also matters as source terms), under free-falling observers' local Lorentz frame \cite{2011PhRvD..84l4014N}, reads \cite{2013AmJPh..81..575P}
\bea
\nabla \cdot \mathcal{E} = 0\,, &\quad& \nabla\times \mathcal{E} = -\frac{\partial \mathcal{B}}{\partial t}\,,\notag \\ 
\nabla \cdot \mathcal{B} = 0\,, &\quad& \nabla\times \mathcal{B} = \frac{1}{c^2}\frac{\partial \mathcal{E}}{\partial t}\,.\label{eq:GEMEvo} 
\eea
When doing an order of magnitude analysis, we can obtain from Eq.~\eqref{eq:GEMEvo} that 
\bea \label{eq:Best}
\|\mathcal{B}\| \sim \frac{\| \mathcal{E}\|}{c} \sim \frac{| \Phi |}{c L^2_{\rm struc}} \sim \frac{G M_{\rm struc}}{c L^3_{\rm struc}} \,,
\eea
where $\|\cdot \|$ denotes a convenient tensor norm, such as Frobenious, and $L_{\rm struc}$ ($M_{\rm struc}$) is the length (mass) scale of cosmic structures relevant for the type of magnetic field we are considering, e.g., it would be the typical length (mass) of galaxy clusters if it is the field in the intergalactic space that we are interested in. Note, it is because the quantities involved vary on such large length scales, that the induced magnetic field ends up possessing such great coherence lengths. 

Let $j={\rm proton}$, and take $M_{\rm struc} \sim \mathcal{O}(10^{15}){\rm M}_{\odot}$,  $L_{\rm struc} \sim \mathcal{O}(1){\rm Mpc}$ as relevant for galaxy clusters, as well $|{\bf v}_p| \sim L_{\rm struc}/T_{\rm struc}$ (regarding $T_{\rm struc}$ also as the temporal duration of the magnetogenesis process, then it cancels out in the final estimate; nevertheless, it might be interesting to note that, e.g., if $T_{\rm struc}\sim 1$Gyr, $|{\bf v}_p|\sim 1000$km/s), we end up with, after substituting Eq.~\eqref{eq:Best} into Eq.~\eqref{eq:genesisorder}, that $|{\bf B}|$ at the end of the gravity-driven magnetogenesis process, becoming of the order 
\bea
|{\bf B}| \sim \mathcal{O}\left(10^{-22}\right){\rm G}\,.
\eea
This value is below the observationally set lower bound mentioned in Sec.~\ref{sec:Intro}, so it can only serve as the seed for further dynamo actions. It is larger than the value at Mpc scales achievable by, e.g., Harrison mechanism \cite{harrison1970generation,2005PhRvD..71d3502M}, and is just above the commonly quoted \cite{2004IJMPD..13..391G,2005PhR...417....1B,2014JCAP...10..056C} lower bound of $\sim \mathcal{O}(10^{-23})$G, on the seed strength required to reproduce the observed $\mu$G galactic magnetic field by dynamo effects [see also \cite{1999PhRvD..60b1301D} though for a much more relaxed bound at $\sim \mathcal{O}(10^{-30})$G].

\section{Conclusion}
In this very brief note, we have attempted an estimate of the magnetic field strength that could possibly be achievable by gravitomagnetism acting upon reionized cosmic plasma. The derivations are crude, enlisting many simplifying assumptions, and the input ingredients to the numerical guesstimate are also unrefined. The predicted field strength could easily swing in either direction by orders of magnitude. Nevertheless, the fact that such a small number of relevant scales combine dimensionally correctly into a  magnetic field strength that falls close to where the seed for dynamos needs to be, is at least intriguing numerologically, thus could be of interest for further investigation, perhaps via numerical simulations that treat gravity relativistically.

Through this example, we also wish to draw attention to the importance of the dynamics intrinsic to the gravitational sector of the Universe. As a possible further utility, we note that the Weyl curvature tensor could affect matter distributions in a way that is attractive on some directions but repelling on others, as gravitational waves impinging on the proverbial ring of free-floating masses would attest. Consequently, a strong contribution from such an influence could possibly aid galactic outflows. 

\acknowledgements
This work is supported by the National Natural Science Foundation of China grants 12073005 and 12021003.

%\bibliography{../../References}
\bibliography{Magnetogenesis.bbl}

\end{document}